\documentclass[twocolumn,prl,preprintnumbers,amsmath,amssymb]{revtex4}

\usepackage{graphicx}
\usepackage{dcolumn}
\usepackage{bm}
\usepackage{feynmp}
\usepackage{epsfig}
\newcommand{\Vzzh}{\ensuremath{V_\mathrm{\scriptscriptstyle ZZH}}}
\newcommand{\Vwwh}{\ensuremath{V_\mathrm{\scriptscriptstyle WWH}}}
\newcommand{\VAwwh}{\ensuremath{V^{A}_\mathrm{\scriptscriptstyle WWH}}}
\newcommand{\gw}{\ensuremath{g_\mathrm{\scriptscriptstyle W}}}
\newcommand{\Mw}{\ensuremath{M_\mathrm{\scriptscriptstyle W}}}
\newcommand{\thetaw}{\ensuremath{\theta_\mathrm{\scriptscriptstyle W}}}

\newcommand{\as}{\ensuremath{\alpha_\mathrm{s}}}
\newcommand{\aw}{\ensuremath{\alpha_\mathrm{\scriptscriptstyle W}}}

\begin{document}

\preprint{Cavendish-HEP-06/26}

\title{QCD and Electroweak Interference in Higgs production by Gauge Boson Fusion}

\author{Jeppe R.~Andersen}
\email{andersen@hep.phy.cam.ac.uk}
\author{Jennifer M.~Smillie}%
\email{smillie@hep.phy.cam.ac.uk} 
\affiliation{%
Cavendish Laboratory, University of
  Cambridge\\JJ Thomson Avenue\\CB3 0HE\\Cambridge, UK
}%

\date{\today}

\begin{abstract}
  We explicitly calculate the contribution to Higgs production at the LHC
  from the interference between gluon fusion and Weak Vector Boson fusion, and
  compare it to the pure QCD and pure Electro-weak result. While the effect
  is small at tree level, we speculate it will be significantly enhanced by
  loop effects.
\end{abstract}

\pacs{Valid PACS appear here}
\maketitle

\section{Introduction}
One of the main tasks for the experimental and theoretical programme in
connection with the CERN LHC is to determine whether the breaking of the
electro-weak symmetry is due to the Higgs boson of the Standard Model. While
the production of a Standard Model Higgs boson at the LHC is dominated by the
process $gg\to H$, gluon fusion mediated through a top quark loop, the
exact dynamics of the symmetry breaking can be more directly extracted by
studying the production of the Higgs in addition to two jets. This process
receives contributions from channels of both
electro-weak\cite{Cahn:1983ip,Dicus:1985zg,Altarelli:1987ue} and QCD
origin\cite{Kauffman:1996ix,DelDuca:2001fn}. Calculations of the
$\mathcal{O}(\as)$ corrections to the electro-weak channel indicate that the
higher order effects are very small, and since the QCD channel can be
efficiently suppressed, it would seem that the Higgs coupling to electro-weak
bosons can be cleanly studied.

The purpose of this letter is to explore mechanisms which could reduce the
purity of the extraction of this $HVV$ coupling from the $Hjj$ process. We
will further highlight the generality of one such observed mechanism.

\section{Higgs + 2 jets in the Standard Model}
\label{sec:higgs-+-2}
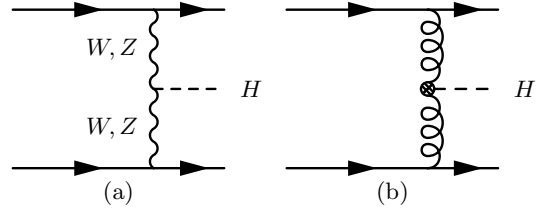
\begin{figure}
  \centering
  \begin{tabular}{cc}
\parbox[c][2.4cm][c]{3.5cm}{\begin{fmffile}{WBF}
    \begin{fmfgraph*}(80,60)
      \fmfpen{thin} \fmfstraight \fmfleft{i1,i2,i3} \fmfright{o1,o2,o3}
      \fmf{phantom}{i1,vl1,vl2,o1} \fmf{phantom}{i3,vu1,vu2,o3} \fmffreeze
      \fmf{boson,label=$W,,Z$,label.side=left}{vl2,vi,vu2}
      \fmffreeze
      \fmf{fermion}{i1,vl2,o1}
      \fmf{fermion}{i3,vu2,o3}
      \fmf{dashes}{vi,o2}
      \fmflabel{$H$}{o2}
    \end{fmfgraph*}
  \end{fmffile}}
  &
  \parbox[c][2.4cm][c]{3.5cm}{  \begin{fmffile}{GFP}
    \begin{fmfgraph*}(80,60)
      \fmfpen{thin} \fmfstraight \fmfleft{i1,i2,i3} \fmfright{o1,o2,o3}
      \fmf{phantom}{i1,vl1,vl2,o1} \fmf{phantom}{i3,vu1,vu2,o3} \fmffreeze
      \fmf{gluon}{vl2,vi,vu2}
      \fmffreeze
      \fmf{fermion}{i1,vl2,o1}
      \fmf{fermion}{i3,vu2,o3}
      \fmf{dashes}{vi,o2}
      \fmflabel{$H$}{o2}
      \fmfv{decor.shape=circle,decor.size=5,decor.filled=hatched}{vi}
    \end{fmfgraph*}
  \end{fmffile}}\\
(a)&(b)
\end{tabular}
\caption{$Hjj$ production at tree level by weak boson fusion and gluon fusion
  (where the effective vertex from Eq.~\eqref{eq:GGHcoupling} is used).}
  \label{fig:WBFandGF}
\end{figure}

We will consider the production of a Standard Model Higgs with a mass between
115~GeV and 200~GeV, in which case the gluon-gluon-Higgs fusion through a top
loop can be described accurately by an effective vertex of the
form\cite{Dawson:1990zj,Djouadi:1991tk}
\begin{align}
  \label{eq:GGHcoupling}
  V(p_a^\mu,k_b^\nu)=-i\ \frac{\as}{3\pi v}\ \delta^{ab}\ (g^{\mu\nu}\ p\cdot k\ -\ p^\nu k^\mu),
\end{align}
where $v$ is the vacuum expectation value of the Higgs field, and $a,b$
denote the colour index of the gluons. Within this approximation, a possible
tree level process for $Hjj$ production is shown in
Fig.~\ref{fig:WBFandGF}(b).  This process is considered a pollutant in the
study of the dynamics of the electro-weak symmetry breaking through the
coupling
\begin{align}
  \label{eq:HWWcoupling}
  \Vwwh(p^\mu,k^\nu)\ =\Vzzh(p^\mu,k^\nu)\cos^2\thetaw=\gw \Mw g^{\mu\nu},
\end{align}
which contributes to $Hjj$--production at tree level through the diagram in
Fig.~\ref{fig:WBFandGF}(a). 
Higher order corrections from QCD are considered
separately\cite{Djouadi:2005gi} for the two processes, and have been
calculated for the gluon fusion process to order $\as^{5}$
\cite{DelDuca:2004wt,Dixon:2004za,Badger:2004ty,Ellis:2005qe,Campbell:2006xx}
in the limit of infinite top mass, and to order $\aw^3\as$ for weak boson
fusion (WBF)\cite{Han:1992hr}. Since higher order corrections to WBF are found to
be small (on the order of $3\%-10\%$ in the
relevant kinematic region \cite{Figy:2003nv,Berger:2004pc}) and the
gluon fusion channel can be suppressed to the same level, it even becomes
feasible to study possible anomalous couplings of the Higgs field to the weak
bosons of the form~\cite{Hankele:2006ma}
\begin{align}
  \begin{split}
    \label{eq:anomalous_couplings}
    &\VAwwh(p^\mu,k^\nu)\ =\ \gw \Mw \big(a_1(p,k) g^{\mu\nu}\\&+\
    a_2(p,k)[g^{\mu\nu}\ p\cdot k
    -p^\nu k^\mu]
    +\
    a_3(p,k)\varepsilon^{\mu\nu\rho\sigma}q_{1\rho}q_{2\sigma}\big),
  \end{split}
\end{align}
where $a_1=1,a_2=a_3=0$ corresponds to the Standard Model, and any
deviation from this is anomalous.

The high purity of the $Hjj$ sample in terms of WBF is obtained by imposing
the following cuts, which are used in our numerical studies
\begin{align}
  \begin{split}
    \label{eq:cuts}
    &R_{jj}>0.6\quad
    \eta_{j_1}\cdot\eta_{j_2}<0\quad \left|\eta_{j_1}-\eta_{j_2}\right|>4.2\\
    &p_{{\scriptscriptstyle T},j}>20\ \mathrm{GeV}\quad |\eta_j|\le 5\quad
    s_{jj}>\left(600\ \mathrm{GeV}\right)^2.
  \end{split}
\end{align}

In this setting it is worth considering all standard model contributions
which could mimic an anomalous coupling and destroy the pure extraction of the
VVH vertex.

One such contribution, which has been ignored in the literature, is the
interference between the QCD and electro-weak generated Higgs production.
Usually (e.g.~in the effective structure function approach), all
contributions requiring a $t\leftrightarrow u$-channel crossing at tree level
are neglected, since such effects only contribute to channels of identical
quarks, and furthermore suffer a kinematical suppression arising from the
crossing. For the $ZZH$ fusion process, the channels with identical quarks
contribute a third of the cross section, but the crossed term suffers the
suppression of a further 4 orders of magnitude due to the kinematical effects
in the crossing.  It is therefore valid to ignore such contributions.  Within
this approximation, there is also no interference between diagrams of colour
singlet and colour octet exchange, and therefore even the tree-level effects
discussed in this letter are ignored.

However, by calculating explicitly the interference between the QCD and
electro-weak generated Higgs production, we find that the
$(\aw\sqrt{\aw}\as^2)$--contribution from this crossed term is 15 times
larger than the neglected contribution from $ZZ\mbox{x}ZZ$--interference,
even though it suffers the same suppression effects from the requirement of a
$t\leftrightarrow u$--crossing and identical quarks.

While this may seem like a small effect, it is about 1\% of the quark
initiated pure gluon fusion channel, and about 5\%-10\% of the NLO QCD
corrections to the quark initiated Z fusion channel reported in
Ref.\cite{Figy:2003nv,Berger:2004pc}. The kinematical dependence of the
interference term is obviously different from either of the pure electro-weak
and the pure QCD terms, and so the applied cuts do not suppress this term to
the same extent as the pure QCD one. In fact, the relative impact of the tree
level interference is increased by raising the cut on the transverse momentum
of the jets.

More importantly, the Lorentz structure of the effective $ggH$--coupling
could be mistaken for an anomalous contribution to the $ZZH$--coupling, since
such contributions are not present in the available NLO calculations of this
process. 

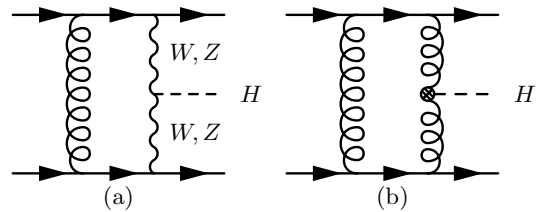
\begin{figure}
  \centering
  \begin{tabular}{cc}
    \parbox[c][2.4cm][c]{3.5cm}{\begin{fmffile}{NLOWBF}
    \begin{fmfgraph*}(80,60)
      \fmfpen{thin} \fmfstraight \fmfleft{i1,i2,i3} \fmfright{o1,o2,o3}
      \fmf{phantom}{i1,vl1,vl2,o1} \fmf{phantom}{i3,vu1,vu2,o3} \fmffreeze
      \fmf{boson,label=$W,,Z$,label.side=right}{vl2,vi,vu2}
      \fmffreeze
      \fmf{gluon}{vu1,vl1}
      \fmf{fermion}{i1,vl1,vl2,o1}
      \fmf{fermion}{i3,vu1,vu2,o3}
      \fmf{dashes}{vi,o2}
      \fmflabel{$H$}{o2}
    \end{fmfgraph*}
  \end{fmffile}} &
  \parbox[c][2.4cm][c]{3.5cm}{\begin{fmffile}{NLOGFP}
    \begin{fmfgraph*}(80,60)
      \fmfpen{thin} \fmfstraight \fmfleft{i1,i2,i3} \fmfright{o1,o2,o3}
      \fmf{phantom}{i1,vl1,vl2,o1} \fmf{phantom}{i3,vu1,vu2,o3} \fmffreeze
      \fmf{gluon}{vl2,vi,vu2}
      \fmffreeze
      \fmf{fermion}{i1,vl1,vl2,o1}
      \fmf{fermion}{i3,vu1,vu2,o3}
      \fmf{dashes}{vi,o2}
      \fmf{gluon}{vu1,vl1}
      \fmflabel{$H$}{o2}
      \fmfv{decor.shape=circle,decor.size=5,decor.filled=hatched}{vi}
    \end{fmfgraph*}
  \end{fmffile}}\\
(a)&(b)
\end{tabular}
\caption{Diagrams contributing to QCD and electro-weak interference terms of order
  $\aw\sqrt{\aw}\as^3$ in the uncrossed channel for all assignments of quark
  flavours.}
  \label{fig:NLOQCDWBF}
\end{figure}

Most importantly though, higher order QCD corrections, which take into
account the exchange of a gluon between the quark lines as depicted in
Fig.~\ref{fig:NLOQCDWBF}, will remove the requirement of a $t\leftrightarrow
u$-crossing in the interference between the QCD and electro-weak processes,
and permit interference effects in all channels. The indicated one-loop
process is the leading order mechanism for processes not suppressed by
crossing, and for all processes involving non-identical quarks, and quarks of
different helicity configurations. The one-loop process is therefore not a
higher order correction to the tree-level process reported above, and the
size of the one-loop interference term is not indicated by the size of the
calculated tree-level interference term. Rather, the size of the contribution
from one-loop interference should be comparable to the size of the one-loop
(in this case NLO) corrections to the weak boson fusion and gluon fusion
processes reported in Ref.\cite{Figy:2003nv,Berger:2004pc,Campbell:2006xx}.
This effect could then have impact beyond the study of possible anomalous
couplings.

\begin{figure}
  \centering
  \epsfig{width=\columnwidth,file=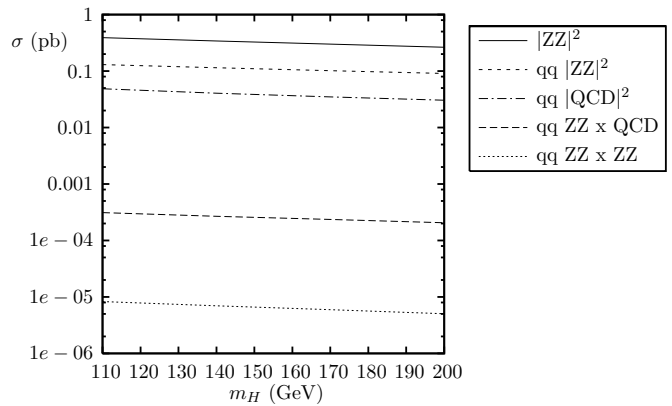}
  \caption{The contribution to $Hjj$ within the cuts of Eq.~(\ref{eq:cuts})
    from various tree level processes as a function of the mass of the Higgs
    boson. ZZxQCD and ZZxZZ denote $t\leftrightarrow u$-channel
    interference terms.}
  \label{fig:H2jet}
\end{figure}

\section{Summary and Conclusions}
\label{sec:summary-conclusions}
We have calculated the interference between colour octet gluon fusion and
colour singlet weak boson fusion channels in $Hjj$--production at tree level.
The interference term has a different kinematic dependence to other
contributions previously investigated and is small (but still an order of
magnitude larger than the $ZZ$ interference term) only due to the tree level
requirement of $t\leftrightarrow u$-channel crossing, which will be lifted at
higher orders and could lead to a very significant interference effect. We
speculate that the one-loop corrections of order $\as^3\aw\sqrt{\aw}$ could
be as important as the one-loop corrections of order $\as\aw^3$ and $\as^5$
already calculated in the literature.

It should be noted that the ideas of interference effects between
electro-weak and QCD generated processes discussed in this letter obviously
are not confined to Higgs masses in the studied range, nor indeed Higgs
production itself. Similar effects were reported in Ref.\cite{Moretti:2005ut}.

\begin{acknowledgments}
  We would like to acknowledge discussions with Alexander Sherstnev on the
  efficient use of CompHEP\cite{Pukhov:1999gg,Boos:2004kh}.
\end{acknowledgments}

\bibliography{Interference}

\end{document}